\documentclass[twocolumn,showpacs,amsmath,amssymb, superscriptaddress]{revtex4}

\usepackage{amsmath}
\usepackage{amssymb}
\usepackage{amsfonts}

\usepackage{graphicx}
\usepackage{float}
\usepackage{color}

\usepackage{amsthm}

\DeclareMathOperator{\Tr}{Tr}

\DeclareMathOperator{\Span}{span}

\DeclareMathOperator{\U}{\it U}
\DeclareMathOperator{\SU}{\it SU}
\DeclareMathOperator{\su}{\mathfrak s\mathfrak u}

\newcommand{\paren}[1]{{\left( #1 \right)}}

\newcommand{\bracket}[1]{{\left[ #1 \right]}}
\newcommand{\angl}[1]{{\left\langle #1 \right\rangle}}

\newcommand{\pder}[2]{{\partial{#1}\over\partial{#2}}}

\newcommand{\oder}[2]{{{d{#1}}\over{d{#2}}}}

\makeatletter
\def\loweq@align#1#2{\lower.6ex\vbox{\baselineskip\z@skip\lineskip\z@
    \ialign{$\m@th#1\hfil##\hfil$\crcr#2\crcr=\crcr}}}
\def\lowsim@align#1#2{\lower.6ex\vbox{\baselineskip\z@skip\lineskip\z@
    \ialign{$\m@th#1\hfil##\hfil$\crcr#2\crcr\sim\crcr}}}
\def\geqq{\mathrel{\mathpalette\loweq@align >}}
\def\leqq{\mathrel{\mathpalette\loweq@align <}}
\def\grsim{\mathrel{\mathpalette\lowsim@align >}}
\def\lesssim{\mathrel{\mathpalette\lowsim@align <}}
\def\gsim{\mathrel{\mathpalette\lowsim@align >}}
\def\lsim{\mathrel{\mathpalette\lowsim@align <}}
\makeatother
\newcommand{\grless} 
{ {\, \raise-.24em\hbox{$<$} \hspace{-0.8em} \raise.31em\hbox{$>$}\, } }
\newcommand{\lessgr} 
{ {\, \raise-.24em\hbox{$>$} \hspace{-0.8em} \raise.31em\hbox{$<$}\, } }
\newfont{\bg}{cmr10 scaled\magstep4}                    
\newcommand{\bigzerou}{\smash{\lower1.7ex\hbox{\bg 0}}}

\newcommand{\nn}{\nonumber \\ }

\newcommand{\R}{{\mathbb R}}

\newcommand{\TT}{{\cal T}}
\newcommand{\crl}[1]{[-\infty,\infty]}

\newcommand{\ket}[1]{|{#1}\rangle}
\newcommand{\kb}[1]{|{#1}\rangle\!\langle{#1}|}

\newcommand{\bra}[1]{\langle{#1}|}

\newcommand{\Ref}[1]{(\ref{#1})}

\newcommand{\wt}{\widetilde}

\newcommand{\da}[1]{#1^\dag}

\newcommand{\pa}[1]{\sigma_{#1}}

\newcommand{\alg}[1]{{\mathfrak #1}}

\newcommand{\av}[1]{\langle#1\rangle}

\newcommand{\vv}[1]{{\boldsymbol{#1}}}
\newcommand{\g}{{\alg g}}

\begin{document}

\title{Time Optimal Unitary Operations}
  \author{Alberto Carlini}
 \email{carlini@th.phys.titech.ac.jp}
 \affiliation{Department of Physics, Tokyo Institute of
 Technology, Tokyo, Japan}
\author{Akio Hosoya}
 \email{ahosoya@th.phys.titech.ac.jp}
 \affiliation{Department of Physics, Tokyo Institute of
 Technology, Tokyo, Japan}
\author{Tatsuhiko Koike}
 \email{koike@phys.keio.ac.jp}
 \affiliation{Department of Physics, Keio University, Yokohama, Japan}
\author{Yosuke Okudaira}
 \email{okudaira@th.phys.titech.ac.jp}
 \affiliation{Department of Physics, Tokyo Institute of
 Technology, Tokyo, Japan}

\date{January 15, 2007}
\begin{abstract}
Extending our previous work on time optimal quantum state evolution
[A. Carlini, A. Hosoya, T. Koike and Y. Okudaira,
 Phys. Rev. Lett. {\bf 96}, 060503 (2006)],
we formulate a variational principle for finding the time optimal realization 
of a target unitary operation, when the available Hamiltonians are subject to certain 
constraints dictated either by experimental or by theoretical conditions.
Since the time optimal unitary evolutions do not depend on the input quantum state
this is of more direct relevance to quantum computation. 
We explicitly illustrate our method by considering the case of a two-qubit system 
self-interacting via an anisotropic Heisenberg Hamiltonian and by deriving the time 
optimal unitary evolution for  three examples of target quantum gates, namely
the swap of qubits, the quantum Fourier transform and the entangler gate.
We also briefly discuss the case in which certain unitary operations
take negligible time.  
\end{abstract}  


\maketitle

\section{Introduction}
Time optimal quantum computation is attracting a growing attention \cite{khaneja,zhang,schulte,boscain} besides the
more conventional concept of optimality in terms of gate complexity, i.e. the number of elementary gates used in a quantum circuit \cite{chuangnielsen}. The minimization of physical time to achieve a given unitary 
transformation is relevant for the design of fast elementary gates.
It also provides a physical ground to describe the complexity of quantum algorithms, whereas gate complexity should
be regarded as a more abstract concept in which physics is implicit. 
Works relevant to the former subject can be found, e.g., in 
\cite{khaneja} and \cite{zhang}, which discuss the time 
optimal generation of unitary operations for a small number of
qubits using a Cartan decomposition scheme and assuming 
that one-qubit operations can be performed arbitrarily fast.
An adiabatic solution to the optimal control problem in holonomic quantum
computation was given in \cite{tanimura}, while 
Schulte-Herbr\"uggen et al. \cite{schulte} numerically obtained  
improved upper bounds on the time complexity of certain quantum gates.
The present authors \cite{CarHosKoiOku06PRL} discussed the quantum brachistochrone for state evolution, i.e.
the problem of finding the time optimal evolution and the optimal Hamiltonian 
of a quantum system for given initial and final states.
Nielsen et al. \cite{nielsen1} proposed a criterion for optimal
quantum computation in terms of a certain 
geometry in Hamiltonian space, and showed in \cite{nielsen2} that
the quantum gate complexity is 
related to optimal control cost problems.
Khaneja et al.~\cite{khanejanew} suggested a geometrical method for
the efficient synthesis of the controlled-{\scshape not} gate between
two qubits with a special Hamiltonian. 

In the standard quantum computation paradigm a whole algorithm may be
reduced to a sequence of unitary transformations between intermediate
states and a final measurement to read the result.
In this paper we address the time optimality of each unitary transformation,
i.e., each subroutine. 
An example is the discrete Fourier transform in Shor's algorithm
for factorization.

In our previous work \cite{CarHosKoiOku06PRL}, the quantum
brachistochrone was formulated as an action principle 
for the quantum state in the complex projective space endowed with the
Fubini-Study metric, and the Hamiltonian subject to certain constraints. 
We obtained the time optimal state evolution and the optimal
Hamiltonian by solving the Euler-Lagrange equations. 
In the present work we extend the methods used in \cite{CarHosKoiOku06PRL} and we
describe the general framework for finding the time optimal
realization of a given unitary operation.
Roughly speaking, we replace the projective space representing 
quantum state vectors with the space of unitary operators.  
While the optimality in the previous work depends on the initial state, it does not in 
the present case so that it is more directly relevant to subroutines in quantum computation,
where the input may be unknown.
This work should be useful not only for designing the efficient
quantum algorithms and devices but also for deepening our insight
into the true origin of the power of quantum computation.

The paper is organized as follows.
In Section~\ref{var} 
we introduce the problem by defining an action principle for the time optimal
realization of unitary operations, under the condition of a Schr\"odinger evolution and
of the existence of a set of constraints for the available Hamiltonians, and we derive the fundamental equations of motion.
We discuss a typical class of the problem in Section~\ref{typical}.
In Section~\ref{ex}
we explicitly show how our formalism works via the example of 
a two-qubit system, which self-interacts by an anisotropic 
Heisenberg Hamiltonian depending
on several control parameters. We derive the time optimal controls
and the optimal time duration required to
generate a {\scshape swap} gate, a {\scshape `qft'} gate and an entangler gate.
A system in which certain operations take negligible time is 
discussed briefly in Section \ref{G/K}. 
Finally, Section~\ref{summary} is devoted to the summary and discussion of our
results.

\section{A variational principle}
\label{var}
Let us consider the problem of 
performing a given unitary operation or a quantum subroutine 
in the shortest time by controlling a certain physical system. 
Mathematically this is a time optimality problem of achieving 
a unitary operator $U_f\in \U(N)$ (modulo overall phases)
by controlling the Hamiltonian $H(t)$ and 
evolving a unitary operator $U(t)$, 
where $H(t)$ and $U(t)$ obey to  the Schr\"odinger
equation.
Note that overall phases are physically irrelevant for quantum evolutions. 
One immediately observes that 
there must be some constraints for $H(t)$, because otherwise one would be able
to realize $U_f$ in an arbitrarily short time simply by
rescaling the Hamiltonian \cite{CarHosKoiOku06PRL}. 
Thus at least the `magnitude' of the Hamiltonian must be bounded. 
Physically this corresponds to the fact that one can afford only a  
finite energy in the experiment. 
Besides this normalization constraint, 
the available Hamiltonians may be subject also to other constraints, which can represent 
either experimental requirements (e.g.,  the specifications of the apparatus in use) 
or theoretical conditions (e.g., allowing no
operations involving three or more qubits). 

We then define the following action for the dynamical variables $U(t)$ and $H(t)$,
\begin{align}
  \label{eq-action}
  S(U,H,\Lambda,\lambda_j) :=&\int dt\left[
  L_T+L_S+L_C\right]
\end{align}
with
\begin{align}
  \label{eq-L}
  &L_T :=
  \sqrt
  {\frac{\av{\tfrac{dU}{dt},(1-P_U)\left (\tfrac{dU}{dt}\right )}}
    {\av{HU,(1-P_U)(HU)}}}, 
  \\
  &L_S :=\av{\Lambda,i \tfrac{dU}{dt}\da U-H},
  \\
  \label{eq-LC}
  &L_C :=
    \sum_j{\lambda_j}f^j(H), 
\end{align}
where we have introduced the Hilbert-Schmidt norm $\angl{A,B}:=\Tr \da AB$ and
the projection $P_U(A):=\tfrac{1}{N}\Tr(A\da U)U$. 
The Hermitian operator $\Lambda(t)$ and the scalars $\lambda_j(t)$ are Lagrange
multipliers. 
The action term $\int L_Tdt$ gives the time duration to be optimized
and corresponds to the action 
$\int [ds/v]$, where $v$ is the velocity of the particle, 
in the classical brachistochrone. 
The metric 
\begin{align}
ds_\text{\tiny$U$}^2=\av{dU,(1-P_U)(dU)}
\label{metric}
\end{align}
is  analogous to the
Fubini-Study metric $ds_\text{\tiny$FS$}^2=\bra{d\psi}(1-\kb{\psi})\ket{d\psi}$ for the
quantum state $\ket{\psi}$ and is invariant under left and right global $U(N)$ 
multiplications. 

The variation of $L_S$ by $\Lambda$ gives the Schr\"odinger equation
\begin{align}
  \label{eq-Sch}
  i\frac{dU}{dt}=HU, \quad\text{or}\quad 
  U(t)={\mathcal  T}e^{-i\int^t_0 Hdt}, 
\end{align}
where $\mathcal T$ is the time ordered product. 
This is similar to the case of the quantum brachistochrone for 
quantum states \cite{CarHosKoiOku06PRL}. 
On the other hand, the variation of $L_C$ by $\lambda_j$ leads to the constraints for $H$,
\begin{align}
  f_j(H)=0. 
\label{constraints}  
\end{align}
If we assume that the constraint functions $f_j(H)$ 
depend only on the traceless part of $H$, i.e.
$\wt H :=(1-P_1)(H)=H-(\Tr H)1/N$,
thanks to the projection $P_U$ in \Ref{eq-L}
the action $S$ is invariant under the $U(1)$ gauge transformation
\begin{align}
U\mapsto e^{i\theta} U, ~H\mapsto H-\tfrac{d\theta}{dt}, ~
  \Lambda\mapsto \Lambda, ~ 
  \lambda_j\mapsto\lambda_j,
  \label{eq-gauge}
  \end{align} 
where $\theta$ is a real function. 
In the following we will consider the time optimal evolution
of operators belonging to the group $\U(N)/\U(1)\simeq \SU(N)$. 
This is natural because overall phases are irrelevant in quantum mechanics.
To present our method in its simplest form, we have 
restricted ourselves to the case where the gauge degree of freedom is $U(1)$. 
However, when there are quantum operations whose time duration is so
short that it can be neglected, we will have a larger gauge group
$K$. Such a case is discussed briefly in Section \ref{G/K}. 

We incidentally note here that, when the Hamiltonian is time independent,
the unitary operator actually evolves along a geodesic with respect to
the metric ${ds}_\text{\tiny$U$}^2$.
This can be easily seen from \Ref{eq-Sch}, which implies
\begin{align}
\tfrac{d}{dt}\left [(1-P_1)\left (\tfrac{dU}{dt}U^{\dagger}\right )\right ]=0,
\label{geodesic}
\end{align}
the same equation as derived from the variation by $U$ of the arclength 
$\int ds_\text{\tiny$U$}$.

Let us now derive the other equations of motion. 
Before taking variations of the action, 
it is convenient to rewrite $L_T$ 
as 
\begin{align}
  \label{eq-LT-tmp}
  L_T=
  \sqrt
  {\frac{\av{\tfrac{dU}{dt} \da U,(1-P_1)(\tfrac{dU}{dt}\da U)}}
    {\av{H,(1-P_1)(H)}}}, 
\end{align}
where we have used the relation $P_U(A)=P_1(A\da U)U$. 
Then the variation of $S$ by $H$ gives 
\begin{align}
  \label{eq-var-S-H}
  - L_T\cdot
  \frac{(1-P_1)(H)}{\av{H,(1-P_1)(H)}}-\Lambda+F=0, 
\end{align}
where 
we have introduced the operator
\begin{align}
  \label{eq-F-def}
  F:=\pder {L_C}H, 
\end{align}
which plays an important role in the following. 
Using \Ref{eq-Sch}, which implies $L_T=1$, 
and recalling that $(1-P_1)(H)=\wt H$, 
one can rewrite \Ref{eq-var-S-H} as 
\begin{align}
  \label{eq-dH}
\Lambda=F-  \frac{\wt H}{\Tr \wt H^2}. 
\end{align}

Let us now take the 
variation of $S$ by $U$. 
We first note that 
\begin{align}
  \label{eq-var-U-tmp}
  \Tr A\delta\paren{\frac{dU}{dt}\da U}=\Tr D[A]U\delta\da U 
\end{align}
for any $A$ 
up to a total time derivative, 
where $D[A]:=\oder At+[A,\oder Ut\da U]$. 
The equation above holds because 
$\delta \da U=-\da U\delta U\da U$ and 
$\tfrac{d\da U}{dt}=-\da U\tfrac{dU}{dt}\da U$. 
Using \Ref{eq-LT-tmp} and \Ref{eq-var-U-tmp}, 
one can easily calculate 
$\delta S/\delta U=0$ to obtain 
\begin{align}
  \label{eq-var-S-U}
  D\left[{L_T}\cdot
      \frac{(1-P_1)(\tfrac{dU}{dt}\da U)}
    {\av{\tfrac{dU}{dt} \da U,(1-P_1)(\tfrac{dU}{dt}\da U)}}
  +i\Lambda \right]=0. 
\end{align}
When the Schr\"odinger equation \Ref{eq-Sch} and
\Ref{eq-dH} for $\Lambda$ hold, 
one finds that the argument of $D$ above is simply $iF$. 
We thus have $D[F]=0$. 
Rewriting this, we obtain the 
{\em quantum brachistochrone equation} 
\begin{align}
  \label{eq-fund}
  i\frac{dF}{dt}= [H, F],
  \quad \text{or} \quad
  F(t)=U(t)F(0)\da U(t). 
\end{align}
This, together with the Schr\"odinger equation \Ref{eq-Sch} and
the constraints \Ref{constraints}, is our fundamental equation \cite{kosloff}.
The quantum brachistochrone equation \Ref{eq-fund} 
seems universal, as it holds also in the case of time optimal evolution of
pure 
\cite{CarHosKoiOku06PRL} and mixed \cite{next} quantum states~%
\cite{oldF}.
In particular, equation \Ref{eq-fund} implies
a simple conservation law, 
\begin{align}
  \label{eq-F^k}
  \Tr F^m=\text{const.}, \quad m=1,2,....
\end{align}

In order to solve the quantum brachistochrone equation \Ref{eq-fund}, 
one should first eliminate the gauge 
freedom \Ref{eq-gauge}. 
The most natural gauge choice is to take $H$ to be traceless, i.e., 
\begin{align}
  \label{eq-tl-gauge}
  H =\wt H. 
\end{align}
This corresponds to choosing the unitary operator $U$ to be an
element of $\SU(N)$. 
Then, for a given operation $U_f$, the procedure to find 
the optimal Hamiltonian $H$ and the optimal time duration
$T$ as follows:

(i) specify the functions $f_j(H)$ which constrain
the range of available Hamiltonians; 

(ii) write down the quantum brachistochrone equation \Ref{eq-fund}; 

(iii) solve \Ref{eq-fund} together with the constraints 
\Ref{constraints} to obtain $H(t)$; 

(iv) integrate the Schr\"odinger equation \Ref{eq-Sch} with $U(0)=1$ 
to get $U(t)$; 

(v) fix the constants in $H(t)$ by imposing the 
condition that $U(T)$ equals $U_f$ modulo a global $U(1)$, i.e., 
 \begin{align}
  U(T)=e^{i\chi}~U_f, 
  \label{targetU}
\end{align}
where $\chi$ is some real number.

In essence, 
we have reduced the problem of 
finding the time optimal unitary evolution, for Hamiltonians subject
to certain constraints, to a set of first-order ordinary differential
equations, which we call the quantum brachistochrone equation.
Such an equation can always be solved in the general
$\U(N)$ case, e.g. numerically.

\section{Typical class of constraints}
\label{typical}
Let us now discuss a typical and important class of constraints. 
We assume that the normalization condition for $H$, 
i.e., the finite energy condition, 
can be written in the form 
\begin{align}
f(H) :=\tfrac{1}{2}(\Tr \wt H^2- N\omega^2)=0,
\label{eq-normH}
\end{align} 
where $\omega$ is a constant. 
Then the constraint part of the Lagrangian can be rewritten as 
\begin{align}
  &L_C=\lambda f(H)+L_C', \label{L_C}
\end{align}
where $\lambda$ is a Lagrange multiplier and
$L_C'$ is the sum of 
the other constraints. 
Therefore, from \Ref{eq-F-def}, \Ref{eq-normH} and \Ref{L_C} we obtain 
\begin{align}
  \label{eq-F}
  F= \lambda \wt{H}+F',
\end{align}  
where $F' :=\pder {L_C'}H$.
Multiplying \Ref{eq-F} by $U$ from the right, using the quantum
brachistochrone equation \Ref{eq-fund}, the Schr\"odinger equation 
\Ref{eq-Sch} and \Ref{eq-tl-gauge}, we have 
$i\lambda\dot U+F'U=UF(0)$. 
By formal integration, we get
\begin{align}
  \label{eq-fec-U}
  U=\bracket{\TT \exp\paren{i\int_0^t\frac{F'dt}{\lambda}}}
  \exp\paren{-iF(0)\int_0^t \frac{dt}{\lambda}}. 
\end{align}

The system becomes particularly simple if 
the constraints for $H$ are, except for the finite energy condition 
\Ref{eq-normH},  
linear and homogeneous in $\wt H$, 
namely, if
\begin{align}\label{Lin}
  L_C'=\Tr \wt HF', 
\end{align}
where $F'=\sum_j\lambda_jg_j$ with $g_j\in\su(N)$,
so that we have:
\begin{equation}\label{tr_gj-H}
\Tr g_j \wt{H}=0.
\end{equation}
Many problems 
in quantum computation or 
quantum control, including the example in the following section, 
fall into this subclass. 
Note that with the assumption \Ref{Lin} $F'$ 
does not depend on the Hamiltonian $H$ explicitly.

We can easily show that $\lambda$ in \Ref{eq-F} is a constant. 
Choosing the gauge \Ref{eq-tl-gauge}, we have
\begin{align}
  0=\Tr \wt H\oder Ft
  =N\omega^2\oder{\lambda}t, 
\end{align}
where the first equality follows from \Ref{eq-fund} and 
the second one from the constraints \Ref{eq-normH} and 
\Ref{tr_gj-H}. Thus $\lambda$ is a constant, which can be chosen equal to one
by a simple rescaling of $F$.
From \Ref{eq-fec-U}, we finally get
\begin{align}
  \label{eq-fec-U-2}
  U=\bracket{\TT\exp\paren{i\int_0^tF'dt}}\exp\paren{-iF(0)t},
\end{align}
while the Hamiltonian $\wt H(t)$ and the Lagrange multipliers $\lambda_j(t)$ are
determined by \Ref{eq-fund}, i.e.
\begin{align}
  \label{eq-fec-fund}
  \oder {\wt H} t+\sum_j\oder{\lambda_j}tg_j=-i\sum_j\lambda_j[\wt H,g_j]. 
\end{align}

\section{Example}
\label{ex}
So far we have developed a general framework for finding the time
optimal Hamiltonian. 
Let us now illustrate our method by solving some specific examples explicitly. 
What we consider is a physical system of two qubits represented by two spins 
interacting via controllable, anisotropic couplings $J_j(t)$ ($j=x, y,
z$) and subject  
to local, controllable magnetic fields $B^a(t)$ ($a=1, 2$) restricted
to the $z$-direction. 
In other words, we choose as an example the following two-qubit Heisenberg Hamiltonian,
\begin{eqnarray}
 H\!\!& :=&\!\!-\sum_jJ_j\sigma_j^{1}\sigma_j^{2}
 +\sum_{a}B^a \sigma_z^{a},
\label{heisenberg}
\end{eqnarray}
where $\sigma_j^{1} :=\sigma_j\otimes 1$,
$\sigma_j^{2} :=1\otimes \sigma_j$ and $\sigma_j$ are the Pauli
operators \cite{flux}.
In the standard computational basis 
labeled as $|00\rangle, |01\rangle, |10\rangle, |11\rangle$, the Hamiltonian \Ref{heisenberg} reads
\begin{eqnarray}
H\!\!&=&\!\!
\left[
\begin{array}{cccc}
\!\! -J_z\!+ \! B_+&\!0 &\!0 &\!-J_-\\
 0& J_z\!+\! B_- & -J_+ & 0\\
 0& -J_+ & J_z\!-\! B_- & 0\\
 -J_-& 0 & 0& -J_z\!-\! B_+\!\!
\label{blockdiagH}
\end{array}
\right],
\end{eqnarray}
where we have introduced $ B_{\pm}(t) :=B^1(t)\pm B^2(t)$ and 
 $J_{\pm}(t) :=J_x(t)\pm J_y(t)$.
 %
By simply reordering the basis states as $|00\rangle, |11\rangle, |01\rangle,
|10\rangle$,
the Hamiltonian can be rewritten as 
$H=H_+\oplus H_-$,
where
$H_{\pm}:=\left[
\begin{array}{cc}
-J_z+B_{\pm}& -J_{\mp}\\
-J_{\mp}& -J_z-B_{\pm}\\
\end{array}\right ]$.
We assume the finite energy condition \Ref{eq-normH}, i.e.
\begin{equation}\label{const-H}
 B_+^2 +B_-^2 +J_+^2+J_-^2+2J_z^2=2\omega^2.
\end{equation}
Then our problem is an example of the
linear homogeneous constraints discussed in the previous
section. 
Namely, 
the form \Ref{heisenberg} of the physical Hamiltonian is guaranteed 
by 
\begin{equation}
 F'=\sum_{j\not = k}\lambda_{jk} \sigma_j^1 \sigma_k^2
  + \sum_{a}\sum_{j=x,y} \lambda_j^{(a)} \sigma_j^a,
\label{fheisenberg}
\end{equation}
where $\lambda_{jk}(t)$ and $\lambda_j^{(a)}(t)$ are Lagrange multipliers.

Our task is to solve the quantum brachistochrone equation
\Ref{eq-fund}, or \Ref{eq-fec-fund}. 
Comparing the coefficients of
the generators of $SU(4)$ on both sides,
we find that the Lagrange multipliers $\lambda_{xy}$ and $\lambda_{yx}$ and 
the coupling $J_z$ are constants.
Furthermore, the control variables $B_{\pm}$ and
$J_{\pm}$ decouple  
from the others and we obtain,
\begin{align}
  B_{\pm}(t)&= B_{0 \pm}\cos 2(\gamma_{\pm} t+\psi_{\pm})\label{sol-B},\\
  J_{\pm}(t)&= \mp B_{0 \mp}\sin 2(\gamma_{\mp} t+\psi_{\mp})\label{sol-J},
\end{align}
where 
$B_{0 \pm}, \psi_{\pm}$ and $\gamma_{\pm}:=\lambda_{xy}\pm \lambda_{yx} $
are constants.

Let us now obtain $U(t)$ by directly solving 
the Schr\"odinger equation \Ref{eq-Sch} instead of using 
\Ref{eq-fec-U-2}. 
Thanks to the block-diagonal form 
of the model Hamiltonian,
the unitary evolution operator is also block diagonal, i.e.
$U=U_+\oplus U_-$. 
Using the Baker-Campbell-Hausdorff formula (see, e.g. \cite{messiah}),
in the permuted computational basis the quantum evolution is described by
the two decoupled equations
\begin{eqnarray}
i\frac{dV_\pm}{dt}&=&(\mp J_z~1\pm\gamma_\pm\sigma_y+B_{0\pm}\sigma_z)V_\pm,
\label{blockU}
\end{eqnarray}
where
$V_{\pm}:= e^{\mp i(\gamma_{\pm}t+\psi_{\pm})\sigma_y}U_{\pm}$. 
Solving \Ref{blockU} and going back to the original (non-permuted)
computational basis, we finally obtain the optimal unitary evolution operator as
\begin{widetext}
\begin{equation}
 U(t)=\left[
\begin{array}{cccc}
 e^{iJ_zt}(\alpha_{0+}+i\alpha_{z+})&0 &0 &e^{iJ_zt}(\alpha_{y+}+i\alpha_{x+})\\
0& e^{-iJ_zt}(\alpha_{0-}+i\alpha_{z-})& e^{-iJ_zt}(\alpha_{y-}+i\alpha_{x-})&0\\
0& e^{-iJ_zt}(-\alpha_{y-}+i\alpha_{x-})& e^{-iJ_zt}(\alpha_{0-}-i\alpha_{z-})&0\\
e^{iJ_zt}(-\alpha_{y+}+i\alpha_{x+}) &0&0& e^{iJ_zt}(\alpha_{0+}-i\alpha_{z+})
\end{array}
\right],
\label{finalU}
\end{equation}
\end{widetext}
where we have chosen $U(0)=1$ and we have introduced the parametrization
\begin{eqnarray}
\alpha_{0 \pm} (t)&:=&\cos\gamma_{\pm} t\cos\Omega_{\pm}t
+ \tfrac{\gamma_{\pm}}{\Omega_{\pm}}\sin\gamma_{\pm}t\sin\Omega_{\pm} t,
\nonumber \\
\alpha_{x\pm}(t)&:=&\pm \tfrac{B_{0\pm}}{\Omega_{\pm}}\sin\Omega_{\pm}t
\sin(\gamma_{\pm}t+2\psi_{\pm}),
\nonumber \\
\alpha_{y\pm}(t)&:=&\pm (\sin\gamma_{\pm}t\cos\Omega_{\pm} t
-\tfrac{\gamma_{\pm}}{\Omega_{\pm}}\cos\gamma_{\pm}t\sin\Omega_{\pm}t ),
\nonumber \\
\alpha_{z\pm}(t)&:=&-\tfrac{B_{0\pm}}{\Omega_{\pm}}\sin\Omega_{\pm}t
\cos(\gamma_{\pm}t+2\psi_{\pm}),
\label{alpha}
\end{eqnarray}
with $\alpha_{0\pm}^2+{\boldsymbol \alpha}_{\pm}^2=1$, and where $\Omega_{\pm}:=\sqrt{B_{0\pm}^2+\gamma_{\pm}^2}$.

The final step is to fix the coefficients $J_z, \alpha_{0\pm}(T)$ and ${\boldsymbol \alpha}_{\pm}(T)$
(i.e. the constants $B_{0\pm}, \gamma_{\pm}$ and $\psi_{\pm}$), the
time duration $T$ and the global phase $\chi$ which realize the target
\Ref{targetU}.
This is done
imposing condition \Ref{targetU} with $U(T)$ expressed via \Ref{finalU} and \Ref{alpha} and $U_f$
represented by the gate that we want to implement in a time optimal way.
We now demonstrate this explicitly by a few simple but interesting examples.

\emph{The {\scshape swap} gate:}
Let us assume that our target $U_f$ is the {\scshape swap} gate 
\begin{eqnarray}
 U_\text{\scshape swap}&:=&
\left[
\begin{array}{cccc}
 1& 0& 0& 0\\
 0& 0& 1& 0\\
 0& 1& 0& 0\\
 0& 0& 0& 1
\end{array}
\right ],
\label{swap}
\end{eqnarray}
which exchanges the states of qubits 1 and 2.
Solving \Ref{targetU} by comparison of the matrix elements of \Ref{finalU} and \Ref{swap} and using
\Ref{alpha}, we obtain the following set of parameters:
$B_{0+}=\gamma_-=0$, $B_{0-}T=\frac{\pi}{2}(1+2p)$, $J_zT=-\frac{\pi}{4}[1-2(p+q)-4(m-n)]$,
$2\psi_-=\frac{\pi}{2}(1+2q)$, and $\chi =-\frac{\pi}{4}[1-2(p+q)+4(m+n)],\label{parswap}$
where $m, n, p$ and $q$ are arbitrary integers and $T$ is still to be determined.
The zero values of $B_{0+}$ and $\gamma_-$, together with \Ref{sol-B} and \Ref{sol-J}, imply that 
$B_{\pm}$ and $J_{\pm}$ are constants and therefore, via \Ref{blockdiagH}, that the optimal Hamiltonian is time independent.
The time optimal duration $T_\text{\scshape swap}$ can then be found by imposing the 
constraint \Ref{const-H}, which  reads
$\left (\frac{4\omega T_{\text{\scshape swap}}}{\pi}\right )^2
=\min_{m,n,p,q}\{2(1+2p)^2+[1-2(p+q)-4(m-n)]^2\}$.
The solutions are $p=0$ and  $q=-2(m-n)$ (or $q=-2(m-n)+1$), which 
lead to $B_1=B_2=0$, $J_x=J_y=2J_z=(-1)^{q+1}\frac{2\omega}{\sqrt{3}}$,
and finally give
$\omega T_\text{\scshape swap}=\tfrac{\sqrt{3}}{4}\pi$ and 
\begin{equation}
 H=(-1)^q\frac{\omega}{\sqrt{3}}\left[
\begin{array}{rrrr}
 1& 0& 0& 0\\
 0&-1& 2& 0\\
 0& 2&-1& 0\\
 0& 0& 0& 1
\end{array}
\right].
\label{Hswap}
\end{equation}
Since the Hamiltonian \Ref{Hswap} is constant, due to \Ref{geodesic} the time optimal evolution is
along a geodesic for the metric $ds_U^2$.
 
\emph{The {\scshape `qft'} gate:}
Suppose now that we want to realize the slightly modified target operation $U_f$ given by
\begin{eqnarray}
 U_{\text{\scshape `qft'}} :=
\left[
\begin{array}{cccc}
 1& 0& 0& 0\\
 0& 0& 1& 0\\
 0& 1& 0& 0\\
 0& 0& 0& i
\end{array}
\right].
\label{swap-like}
\end{eqnarray}  
This gate is important as it is essentially equivalent to performing a quantum Fourier transform ({\scshape qft})
over two qubits, i.e.
$U_{\text{\scshape qft}}=W_1U_{\text{\scshape `qft'}}W_1$, where $W_1$ is the Hadamard transform
acting on qubit 1, i.e. 
$W_1:=\frac{1}{\sqrt{2}}\left[
\begin{array}{rr}
 1& 1\\
 1& -1
\end{array}\right ]\otimes 1$ 
and where the action of the {\scshape qft} on the states of the two-qubit computational basis $\{\ket{x} ~| ~x=0, 1, 2, 3\}$ is given by \cite{chuangnielsen}  
$U_{\text{\scshape qft}} ~:~\ket{x}~\mapsto ~\frac{1}{2}\sum_{y=0}^3e^{\pi i xy/2}\ket{y}$.
The {\scshape qft} is at the core of many quantum algorithms, such as the celebrated Shor's algorithm \cite{shor}
for factoring integers.
If we can assume that the Hadamard transform takes negligible time, 
our methods generate the time optimal Hamiltonian to obtain the target $U_{\text{\scshape qft}}$.
Following steps similar to those for the {\scshape swap} gate, we obtain
the following set of parameters:
 $B_{0+}T=(-1)^r\tfrac{\pi}{4}(1+4r)$,
$B_{0-}T=\tfrac{\pi}{2}(1+2p)$, 
$J_zT=-\tfrac{\pi}{8}[1-4(p+q)-8(m-n)]$, $\gamma_{\pm}=0$, 
$2\psi_+=\pi r$, 
$2\psi_-=\tfrac{\pi}{2}(1+2q)$ and
$\chi =-\tfrac{3\pi}{8}[1-\tfrac{4}{3}(p+q)+\tfrac{8}{3}(m+n)],
\label{parswap-like}$
where again $m, n, p, q, r$ and $s$ are arbitrary integers.
As in the case of the $U_\text{\scshape swap}$ gate, the zero values of the parameters $\gamma_{\pm}$ and eqs. \Ref{sol-B} and \Ref{sol-J} imply 
that $B_{\pm}$ and $J_{\pm}$ are constant and give 
a time independent optimal Hamiltonian $H_\text{\scshape swap}$, and consequently 
a geodesic evolution with respect to $ds_U^2$. 
Imposing the constraint \Ref{const-H} we obtain
$\left (\tfrac{8\omega T}{\pi}\right )^2
=\min_{m,n, p,q,r}\{8(1+2p)^2+[1-4(p+q)-8(m-n)]^2+2(1+4r)^2 \},$
 which is solved by  $p=r=0$ and $q=-2(m-n)$.
This leads to $B_1=B_2=\frac{2\omega}{\sqrt{11}}$,
$J_+=J_-=-\frac{4\omega}{\sqrt{11}}$ and finally gives the optimal time duration 
$\omega T_{\text{\scshape `qft'}}=\tfrac{\sqrt{11}}{8}\pi$ and the optimal Hamiltonian
\begin{equation}
 H=\frac{\omega}{\sqrt{11}}\left[
\begin{array}{rrrr}
 3& 0& 0& 0\\
 0&-1& 4& 0\\
 0& 4&-1& 0\\
 0& 0& 0&-1
\end{array}
\right].
\end{equation}

\emph{The entangler gate:}
As a last example, we want to find the optimal way to generate the entangler gate
\begin{eqnarray}
 U_\text{\scshape ent}&:=&
\left[
\begin{array}{cccc}
 \cos\varphi& 0& 0&\sin\varphi \\
 0& 1& 0& 0\\
 0& 0& 1& 0\\
-\sin\varphi & 0& 0&\cos\varphi 
\end{array}
\right ],
\label{cat}
\end{eqnarray}
where we choose the angle $\varphi\in [0, \pi]$.
This gate, upon acting on the initial state $\ket{00}$, produces the $\varphi$-dependent entangled state $\cos\varphi\ket{00}
-\sin\varphi\ket{11}$.
For example, when $\varphi=3\pi/4, \pi/4$, this allows reaching the maximally entangled Bell states $\ket{\Phi^{\pm}}:=(\ket{00}\pm \ket{11})/\sqrt{2}$.
As usual, comparison of \Ref{finalU} and \Ref{cat} leads to
the following set of parameters:
 $B_{0+}T=(-1)^r\pi\sqrt{p^2-(q-x)^2}$,
$B_{0-}=0$, 
$J_zT=\tfrac{\pi}{2}(m-n)$, 
$\gamma_+T=\pi (q-x)$ and
$\chi =-\tfrac{\pi}{2}(m+n),
\label{parent}$
where $x:=\varphi/\pi$ and, again, $m, n, p, q$ and  $r$ are arbitrary integers.
Imposing the constraint \Ref{const-H} we now obtain
$
\left (\tfrac{2\omega T}{\pi}\right )^2
=\min_{m,n,p,q}\{2[p^2-(q-x)^2] +(m-n)^2\},
$
which is solved by  $m=n$ and $|p|=-q=1$, and leads to $B_1=B_2=
(-1)^p\sqrt{2}\omega\cos (2\gamma_+t+\psi_+)$,
$J_x=J_y=(-1)^p\sqrt{2}\omega\sin (2\gamma_+t+\psi_+)$.
We finally obtain the optimal, $\varphi$-dependent $\omega T_\text{\scshape ent}=\pi\sqrt{x(1-x/2)}$ and 
\begin{equation}
 H(t)=\pm \sqrt{2}\omega\left[
\begin{array}{cccc}
-\cos \mu (t)& 0& 0&\sin \mu (t) \\
 0&0& 0& 0\\
 0& 0&0& 0\\
 \sin \mu (t)& 0& 0&\cos \mu (t) 
\end{array}
\right],
\end{equation}
where $\mu(t)\!:=2(\gamma_+t +\psi_+)$, $\gamma_+(x)\!=\omega (x-1)/\sqrt{x(1-x/2)}$
(see figure 1).
In this case the Hamiltonian is time dependent and, therefore, the time optimal generation of
the entangler gate does not occur along a geodesic.
\begin{figure}[H]
 \begin{center}
  \resizebox{7cm}{!}{\includegraphics{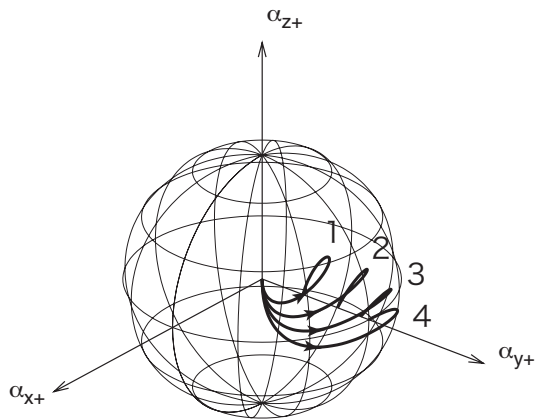}}
  \caption{Time optimal realizations 
    \Ref{finalU} 
    of the entangler gate \Ref{cat} with
    $\varphi=\frac{k\pi}{8}$, $k=1,2,3,4$,  
    from the identity (origin) 
    in the $\vv{\alpha}_+$-space. 
    They are not geodesics. 
  }
 \end{center}
\end{figure}

\section{Case with fast operations}
\label{G/K}
Let us now briefly discuss the quantum brachistochrone 
in the case where certain unitary operations 
have negligible time duration \cite{qit14tk}. 
We assume that such operations (together with the unphysical global
phase transformations) form a subgroup $K$ of the group
$G=\U(N)$ of all the unitary operations. 
We denote the Lie algebras of $G$ and $K$ by $\alg g$ and $\alg k$,
respectively. 
Note that the formulation in Section~\ref{var} can be considered as 
the special case of $K=\U(1)$. 

In order to measure the time duration properly, 
we have to generalize the projection operator $P_U=P_1(A\da U)U$ 
in \Ref{eq-L} 
so that $P_1$ is 
the orthogonal projection to $\alg k$ in $\alg g$, 
namely, 
\begin{align}
  \label{eq-G/K-P}
  P_1(A):=\sum_j\av{g_j,A}g_j, 
\end{align}
where $\{g_j\}$ is an orthonormal basis for $\alg k$. 
Apart from \Ref{eq-G/K-P}, the 
action \Ref{eq-action} and the Lagrangian terms 
\Ref{eq-L}-\Ref{eq-LC} are unchanged. 
Therefore, defining again $\wt H:=(1-P_1)(H)$, we can repeat the same
argument and obtain the same remaining equations 
of Section \ref{var} and \ref{typical}. 
In particular, we obtain the quantum 
brachistochrone equation \Ref{eq-fund} and
we can still follow the procedure (i)-(v) of Section \ref{var}.

The terms $L_T$ and $L_S$ of the Lagrangian are now invariant under
the (in general non-Abelian) gauge transformation 
\begin{align}
  \label{eq-G/K-gauge-trf}
  &U\mapsto kU, \quad
  H\mapsto kH\da k+i\dot k\da k, \nn
  &\Lambda\mapsto k\Lambda\da k, \quad
  \lambda_j\mapsto\lambda_j, 
\end{align}
where $k(t)\in K$. 
Since 
$F$ transforms as $F\mapsto kF\da k$ under \Ref{eq-G/K-gauge-trf}, 
the quantum brachistochrone equation \Ref{eq-fund} is always 
covariant, i.e., unchanged. 
Furthermore, if $L_C$ is invariant under \Ref{eq-G/K-gauge-trf}, 
the constraints are also gauge-invariant. 

One of the systems which is often discussed in quantum computation 
is that of $n$ qubits 
in which the one-qubit operations take negligible time. 
This corresponds to the case $G=\U(2^n)$ and 
$K=\U(1)\otimes\SU(2)^{\otimes n}$. 
Let $\g_j$ be the subspace of $\g$ representing infinitesimal $j$-qubit
operations. 
Namely, $\g_j$ consists of linear combinations of all the 
operators which are products of $j$ Pauli operators and $n-j$ identity
operators: 
\begin{align}
  \g_j&:=
  \Span_\R\{\sigma^{a_1\cdots a_j}_{ l_1\cdots  l_j}; 
  a_1<\cdots<a_j \text{ and } l_j=x,y,z\},
  \nn
  &\qquad j\ge0, 
\end{align}
where $a_m$ represents the $a_m$th qubit for $m=1,\cdots,j$ and 
$\sigma^{a_1\cdots a_j}_{l_1\cdots l_j}$ is a generalization of
$\sigma^a_j$ appearing in \Ref{heisenberg}. 
For example, $\sigma^{13}_{xy}=\pa x\otimes1\otimes\pa
y\otimes1\otimes\cdots\otimes1$. 
Then we have $\alg k=\g_0\oplus\g_1$. 
Moreover, we can write \Ref{eq-G/K-P} explicitly as 
$P_1=\sum_jP^{(j)}$ with 
\begin{align}
P^{(j)}(A):=\sum_{a_1<\cdots<a_j}\sum_{ l_1,\cdots, l_j}
\sigma^{a_1\cdots a_j}_{ l_1\cdots l_j} 
\Tr\paren{\sigma^{a_1\cdots a_j}_{ l_1\cdots l_j}A}
/{2^{j}}. 
\end{align}
Note that each $P^{(j)}$ is the orthogonal projection to $\g_j$ in
$\g$.  

Let us also assume that the infinitesimal operations including 
three or more qubits are not allowed in the Hamiltonian. 
This is the case of the linear homogeneous 
constraints discussed in Section~\ref{typical} with
$F'=\sum_{j=3}^nF'_j$, where
\begin{align}
&F'_j:=\sum_{a_1<\cdots<a_j}\sum_{ l_1,\cdots, l_j}
\lambda^{a_1\cdots a_j}_{ l_1\cdots l_j}
\sigma^{a_1\cdots a_j}_{ l_1\cdots l_j}/{2^j}, 
\end{align}
and $\lambda^{a_1\cdots a_j}_{ l_1\cdots l_j}$ are 
Lagrange multipliers. 
By choosing the gauge $H=\wt H$ 
we have $F'_j\in\g_j$, 
while from the constraints $f_j=0$ with $j\ge3$ 
we get $\wt H\in\g_2$. 
We find that the following commutation relations of the subspaces $\g_j$ 
of the algebra $\g$ hold: 
\begin{align}
  \label{eq-g-comm}
  &[\g_j,\g_k]=\g_{|j-k|+1}\oplus\g_{|j-k|+3}
  \oplus\cdots\oplus\g_{j+k-1}, 
  \quad j,k\ge1,
  \nn
  &[\g_0,\g_j]=0,\quad j\ge0, 
\end{align}
where we understand that $\g_j=0$ for $j>n$. 

In particular, the three-qubit case, $n=3$, turns out to be simple and 
we can 
carry out the procedure in Section~\ref{var} up to (iv) in general
\cite{qit14tk}.  
In fact, by \Ref{eq-g-comm} we have $[\g_2,\g_3]=\g_2$, 
and since $\wt H\in\g_2$ and $F'=F'_3\in\g_3$, 
the quantum brachistochrone equation \Ref{eq-fund} 
decouples into two equations 
\begin{align}
  \dot{\wt H}=-i[\wt H,F'],\quad\dot F'=0.
\end{align}
Thus we have 
$F'(t)=F'(0)$
and $\wt H(t)=e^{iF'(0)t}\wt H(0)e^{-iF'(0)t}$, 
so that we can drop 
the time ordering in \Ref{eq-fec-U-2}, 
and we finally obtain 
\begin{align}
  \label{eq-G/K-U}
  U(t)=e^{iF'(0)t}e^{-i(\wt H(0)+F'(0))t}. 
\end{align}
A similar result was recently found in another setting \cite{dowling}. 
Although the three-qubit system is particularly simple, 
the prescription (v) still remains technically involved.
We postpone its full analysis to a future work.

\section{Summary and discussion}
\label{summary}
We have studied the problem of finding the time optimal evolution of a
unitary operator in $U(N)$ and the corresponding time optimal
Hamiltonian within the context of a variational principle. 
Our main result is an explicit prescription for finding the time
optimal unitary operation.
Once the constraints for the available
Hamiltonians are specified, 
the quantum brachistochrone equation can be immediately written down, 
and then the problem simply reduces to obtaining its solutions. 
Our formulation is general, systematic, and does not rely upon any
restrictive assumptions, e.g., adiabaticity of quantum evolutions. 
We explicitly showed our methods and found 
the optimal Hamiltonian and the optimal time duration for
three important examples of quantum gates acting on two qubits.
The optimal Hamiltonians realizing the {\scshape swap} and {\scshape qft} gates
are time independent and, therefore, the corresponding optimal unitary operators
follow geodesic curves on the $SU(4)$ manifold endowed with the metric $ds_{\text{\tiny$U$}}^2$.
This is not the case for the entangler gate (as expected for generic gates) where the optimal Hamiltonian is time dependent and the time evolution of the corresponding unitary operator is not geodesic.
We also discussed the quantum brachistochrone for unitary
operations in the case where 
there are operations whose time duration is negligible. 

This work is a natural extension of our previous analysis of the time
optimal evolution of quantum states in the projective space. 
The present formulation has direct relevance to quantum computation,
since it gives the optimal realization of subroutines for unknown input
states, e.g. the discrete Fourier transform.
On the other hand, the quantum brachistochrone for state
evolution in \cite{CarHosKoiOku06PRL} may be viewed as a quantum computation for known initial
states, e.g. the transition from $|00..0\rangle$ to a certain entangled state
\cite{shor} in Shor's factorization algorithm.

We should caution the reader that, in order to make the variational principle well defined, the action \Ref{eq-action} should be actually expressed as an integration over a parameter with fixed initial and final values.
Since this does not affect our results, we have omitted these details for simplicity.
Furthermore, we note that, instead of \Ref{eq-L}, any function of $i{\dot U}U^{\dagger}$
and $H$ which becomes constant upon using the Schr\"odinger equation would produce the
same quantum brachistochrone equation \Ref{eq-fund}.
In this sense, the explicit expression of the metric in \Ref{eq-L} does not affect our formulation.
In a related work, the authors of \cite{nielsen1}, \cite{nielsen2} and \cite{nielsen3} rephrased the problem of finding efficient quantum algorithms in terms of the shortest path in a curved geometry.
Their goal was to obtain a bound on the number of gates required to synthesize a given target
unitary operation in terms of a cost function based on a certain metric
in the space of Hamiltonians.
By tailoring the form of such a metric they were able to approximate the
target unitary operation by a circuit of size polynomial in the distance from the identity.
On the other hand, our point of view here is that the time complexity of
an algorithm is of more physical relevance than its gate complexity 
(see also, e.g. \cite{schulte}).
Furthermore, although our result does not depend on the choice of the metric on $U(N)$,
the bi-invariant metric \Ref{metric} is the most natural.
Also note that the simplest isotropic constraint \Ref{eq-normH} 
does not provide any non-trivial bound to the gate complexity.
In our framework the general relationship
between time and gate complexity is still an open issue.

Another point which we would like to emphasize is that, although what we treated here for the 
simplicity of exposition was the case in which the constraints are expressed as equality conditions for the functions $f_j(H)$, there should be no conceptual difficulty in extending our variational methods to the more realistic case when similar constraints are given in terms of inequalities (see. e.g., \cite{khun}).

Finally, we should note that the authors of \cite{khaneja}, by using the Pontryagin maximal
principle, also showed an optimal time dependent Hamiltonian 
as a particular solution to an equation which is
similar to our quantum brachistochrone equation. 
In the two-qubit demonstration of our variational methods, we
have obtained a general solution for the optimal Hamiltonian without
attempting to match it to a prescribed NMR experiment, which was a
main concern in \cite{khaneja}. 
Our formalism also naturally allows for the treatment of the
more general and physical 
situation in which one-qubit local controls require a
non-zero time cost. 

\section*{ACKNOWLEDGEMENTS}
We would like to thank
Professor I. Ohba and Professor H. Nakazato for useful comments.
This research was partially supported by the MEXT of Japan, 
 under grant No. 09640341 (A.H. and T.K.), by the JSPS with
grant L05710 (A.C.) and 
by the COE21 project on `Nanometer-scale Quantum Physics' at Tokyo
 Institute of Technology 
(A.H. and Y.O.).

\bibliographystyle{alpha}

\begin{thebibliography}{1}
\bibitem{oldF}
The definition of $F$ in 
  \cite{CarHosKoiOku06PRL}, i.e. Equation (17) there, 
is different from \Ref{eq-F-def} of the present paper. 
However, one can easily show that Equation (16) in \cite{CarHosKoiOku06PRL} 
is, with the help of (15) there and the Schr\"odinger equation, 
equivalent to the quantum brachistochrone equation 
\Ref{eq-fund} in the present paper with the present definition of 
$F$.
\end{thebibliography}

\end{document}